\documentclass[11pt,twoside]{article}
\usepackage{amssymb,amsmath,amscd}
\usepackage{fancyheadings}
\usepackage{bm}

\newfont{\msbmfont}{msbm10 at 12pt}  
\newfont{\msbmifont}{msbm8}  

\usepackage{graphicx}
\usepackage{caption2}
\setcaptionwidth{\linewidth}

\makeatother

\newcommand{\Nazva}[1]{\begin{center}\baselineskip=6.0mm{\Large\textbf{#1}}\end{center}}

\newcommand{\Avtor}[1]{\centerline{\large\textbf{\copyright \ #1}}}

\newcommand{\email}[1]{\begin{center}{\small\texttt{#1}}\end{center}}
\newcommand{\address}[1]{\begin{center}{\small\textit{#1}}\end{center}}

\newcommand{\lit}[3]{\vspace*{0.7mm}\par\noindent\makebox[5.2mm][r]{#1.~}\parbox[t]{159.8mm}{{\textit{#2}}~{#3}}\hspace*{-1.6mm}}

\begin{document}

\hyphenation{сє-фхь т√-яєъ-ыюё-Єш т√-яєъ-ыющ хё-ыш ърц-фр  ърц-ф√щ
ърц-фю-ую ърц-фющ ъюу-фр ъю-Ёрё-яЁх-фх-ых-эшх ьрЄ-Ёш-Ўр ьрЄ-ЁшЎ
ьрЄ-Ёшў-э√ї ьрЄ-Ёш-Ў√ ьрЄ-Ёш-Ўє ьрЄ-Ёшў-эю-ую юс-ырёЄ№ юс-ырё-Єш
яюф-ёЄрт-ы   яЁхф-ёЄрт-ых-эю яю-ью∙№■ Ёрё-ёьюЄ-Ёшь ёшё-Єх-ьр
ёшё-Єх-ьє ёшё-Єх-ьющ ёшё-Єхь ёшё-Єх-ь√ ёшё-Єх-ьр-ьш ёшё-Єх-ьх
ёшё-Єх-ьрь ёъю-ЁюёЄ№■ ёют-яр-фр-■Є ёюё-ЄюшЄ Єюу-фр Єюц-фхёЄ-тхэ-эю
Єюц-фхёЄ-тр єяЁрт-ых-эшх єяЁрт-ы -х-ь√ї єяЁрт-ы -х-ь√х єяЁрт-ы -х-ьр
ўрё-Єш ўрё-Є ь ўшё-ыю ўшё-ыр}


\newcommand{\hypergeom}[5]{\mbox{$
_#1 F_#2\left. \!\! \left( \!\!\!\!
\begin{array}{c}
\multicolumn{1}{c}{\begin{array}{c} #3
\end{array}}\\[1mm]
\multicolumn{1}{c}{\begin{array}{c} #4
\end{array}}\end{array}
\!\!\!\! \right|\displaystyle{#5}\right) $} }
\newcommand{\SO}{SO}\newcommand{\spin}{{\bf Spin}}\newcommand{\Sp}{Sp}\newcommand{\bi}{{\bf i}}
\newcommand{\bj}{{\bf j}}
\newcommand{\bk}{{\bf k}}
\Nazva{FUNCTIONS OF REPRESENTATIONS OF THE CLASS 1 ON THE
HOMOGENEOUS SPACES OF THE DE SITTER GROUP}
\Avtor{V. V. Varlamov } \email{root@varlamov.kemerovo.su}
\address{Siberia State University of Industry, Novokuznetsk}
\medskip
A starting point of this research is an analogue between universal
coverings of the Lorentz and de Sitter groups, which was first
established by Takahashi [1] (see also the work of Str\"{o}m [2]).
Namely, the universal covering of $\SO_0(1,4)$ is
$\spin_+(1,4)\simeq\Sp(1,1)$ and the spinor group $\spin_+(1,4)$ is
described in terms of $2\times 2$ quaternionic matrices. Spherical
functions on the group $\SO_0(1,4)$ are understood as functions of
representations of the class 1 realized on the homogeneous spaces of
$\SO_0(1,4)$. A list of homogeneous spaces of $\SO_0(1,4)$,
including symmetric Riemannian and non-Riemannian spaces, consists
of the group manifold $\mathfrak{S}_{10}$ of $\SO_0(1,4)$,
two-dimensional quaternion sphere $S^q_2$, four-dimensional
hyperboloid $H^4\sim\SO_0(1,4)/\SO(4)$, three-dimensional real
sphere $S^3\sim\SO(4)/\SO(3)$ and a two-dimensional real sphere
$S^2\sim\SO(3)/\SO(2)$.

Using the universal covering $\spin_+(1,4)\simeq\Sp(1,1)$ of
$\SO_0(1,4)$, we can write a first Casimir operator $F$ on the group
manifold $\mathfrak{S}_{10}$,
\[
-F=\frac{\partial^2}{\partial{\theta^q}^2}+
\cot\theta^q\frac{\partial}{\partial\theta^q}+
\frac{1}{\sin^2\theta^q}\frac{\partial^2}{\partial{\varphi^q}^2}-
\frac{2\cos\theta^q}{\sin^2\theta^q}
\frac{\partial^2}{\partial\varphi^q\partial\psi^q_1}+
\cot^2\theta^q\frac{\partial^2}{\partial{\psi^q_1}^2}+
\frac{\partial^2}{\partial{\psi^q}^2},\hspace{2.5cm}\text{(1)}
\]
where
\[
{\renewcommand{\arraystretch}{1.55}
\begin{array}{ccl}
\dfrac{\partial}{\partial\theta^q}&=&\dfrac{\partial}{\partial\theta}+
\dfrac{\partial}{\partial\phi}+\bi\dfrac{\partial}{\partial\tau},\\
\dfrac{\partial}{\partial\varphi^q}&=&\dfrac{\partial}{\partial\varphi}+
\bi\dfrac{\partial}{\partial\epsilon}+\bj\dfrac{\partial}{\partial\varsigma},\\
\dfrac{\partial}{\partial\psi^q}&=&\dfrac{\partial}{\partial\psi}+
\bi\dfrac{\partial}{\partial\varepsilon}+\bi\dfrac{\partial}{\partial\omega}+
\bk\dfrac{\partial}{\partial\chi},\\
\dfrac{\partial}{\partial\psi^q_1}&=&\dfrac{\partial}{\partial\psi}+
\bi\dfrac{\partial}{\partial\varepsilon}+
\bk\dfrac{\partial}{\partial\chi}.
\end{array}
\quad
\begin{array}{ccl}
\dfrac{\partial}{\partial\dot{\theta}^q}&=&\dfrac{\partial}{\partial\theta}-
\dfrac{\partial}{\partial\phi}-\bi\dfrac{\partial}{\partial\tau},\\
\dfrac{\partial}{\partial\dot{\varphi}^q}&=&\dfrac{\partial}{\partial\varphi}-
\bi\dfrac{\partial}{\partial\epsilon}-\bj\dfrac{\partial}{\partial\varsigma},\\
\dfrac{\partial}{\partial\dot{\psi}^q}&=&\dfrac{\partial}{\partial\psi}-
\bi\dfrac{\partial}{\partial\varepsilon}-\bi\dfrac{\partial}{\partial\omega}-
\bk\dfrac{\partial}{\partial\chi},\\
\dfrac{\partial}{\partial\dot{\psi}^q_1}&=&\dfrac{\partial}{\partial\psi}-
\bi\dfrac{\partial}{\partial\varepsilon}-
\bk\dfrac{\partial}{\partial\chi}.
\end{array}}
\]
Here, $\psi$, $\varphi$, $\theta$, $\phi$, $\varsigma$, $\chi$,
$\tau$, $\epsilon$, $\varepsilon$, $\omega$ are Euler angles of
$\Sp(1,1)$, $\theta^q=\theta+\phi-\bi\tau$,
$\varphi^q=\varphi-\bi\epsilon+\bj\varsigma$,
$\psi^q=\psi-\bi\varepsilon-\bi\omega+\bk\chi$ are quaternion Euler
angles. The second Casimir operator $W$ of $\SO_0(1,4)$ is equal to
zero on the representations of the class 1.

Matrix elements $t^{\sigma}_{mn}(\mathfrak{q})=
\mathfrak{M}^{\sigma}_{mn}(\varphi^q,\theta^q,\psi^q)$ of
irreducible representations of the group $\SO_0(1,4)$ are
eigenfunctions of the operator (1):
\[
\left[-F+\sigma(\sigma+3)\right]\mathfrak{M}^{\sigma}_{mn}(\mathfrak{q})=0,
\hspace{11cm}\text{(2)}
\]
where
\[
\mathfrak{M}^{\sigma}_{mn}(\mathfrak{q})=e^{-\bi(m\varphi^q+n(\psi^q_1-\bi\omega))}
\mathfrak{Z}^{\sigma}_{mn} (\cos\theta^q),\hspace{9.15cm}\text{(3)}
\]
since $\psi^q=\psi^q_1-\bi\omega$. Here,
$\mathfrak{M}^\sigma_{mn}(\mathfrak{q})$ are general matrix elements
of the representations of $\SO_0(1,4)$, and
$\mathfrak{Z}^\sigma_{mn}(\cos\theta^q)$ are {\it hyperspherical
functions}. Substituting the functions (3) into (2) and taking into
account the operator (1), after substitution $z=\cos\theta^q$ we
arrive at the following differential equation:
\[
\left[(1-z^2)\frac{d^2}{dz^2}-2z\frac{d}{dz}-
\frac{m^2+n^2-2mnz}{1-z^2}+\sigma(\sigma+3)\right]
\mathfrak{Z}^{\sigma}_{mn}(z)=0.\hspace{5cm}\text{(4)}
\]
The latter equation has three singular points $-1$, $+1$, $\infty$.
It is a Fuchsian equation. A particular solution of (4) can be
expressed via the hypergeometric function
\begin{multline}\nonumber
\mathfrak{Z}^\sigma_{mn}(\cos\theta^q)=C_1\sin^{|m-n|}\frac{\theta^q}{2}
\cos^{|m+n|}\frac{\theta^q}{2}\times\\
\times\hypergeom{2}{1}{\sigma+3+\frac{1}{2}(|m-n|+|m+n|),-\sigma+\frac{1}{2}(|m-n|+|m+n|)}
{|m-n|+1}{\sin^2\frac{\theta^q}{2}}.\hspace{0.5cm}\text{(5)}
\end{multline}
An explicit form of the functions
$\mathfrak{Z}^\sigma_{mn}(\cos\theta^q)$ can be derived via the
multiple hypergeometric series. Namely, using an addition theorem
for generalized spherical functions [3], we obtain
\begin{multline}\nonumber
\mathfrak{Z}^\sigma_{mn}(\cos\theta^q)=\sqrt{\frac{\Gamma(\sigma+m+1)\Gamma(\sigma-n+1)}
{\Gamma(\sigma-m+1)\Gamma(\sigma+n+1)}}
\cos^{2\sigma}\frac{\theta}{2}\cos^{2\sigma}\frac{\phi}{2}\cosh^{2\sigma}\frac{\tau}{2}\times\\
\sum^\sigma_{k=-\sigma}\sum^\sigma_{t=-\sigma}\bi^{m-k}\tan^{m-t}\frac{\theta}{2}\tan^{t-k}\frac{\phi}{2}
\tanh^{k-n}\frac{\tau}{2}\times\\
\hypergeom{2}{1}{m-\sigma,-t-\sigma}{m-t+1}{-\tan^2\frac{\theta}{2}}
\hypergeom{2}{1}{t-\sigma,-k-\sigma}{t-k+1}{-\tan^2\frac{\phi}{2}}
\hypergeom{2}{1}{k-\sigma,-n-\sigma}{k-n+1}{\tanh^2\frac{\tau}{2}}
\hspace{0.5cm}\text{(6)}
\end{multline}
for $m\geq t,\;t\geq k,\;k\geq n$. In addition to (6) there exist
seven functions $\mathfrak{Z}^\sigma_{mn}(\cos\theta^q)$ for $m\geq
t,\;k\geq t,\;k\geq n$; $t\geq m,\;k\geq t,\;n\geq k$; $t\geq
m,\;t\geq k,\;n\geq k$; $t\geq m,\;k\geq t,\;k\geq n$; $t\geq
m,\;t\geq k,\;k\geq n$; $m\geq t,\;t\geq k,\;n\geq k$; $m\geq
t,\;k\geq t,\;n\geq k$.

Hyperspherical functions for other homogeneous spaces of
$\SO_0(1,4)$ are particular cases of the functions (6). For example,
on the quaternion 2-sphere we have associated functions
$\mathfrak{Z}^m_\sigma(\cos\theta^q)$. Further, the function (6) is
reduced to the Jacobi function $\mathfrak{P}^\sigma_{mn}(\cosh\tau)$
on the hyperboloid $H^4\sim\SO_0(1,4)/\SO(4)$ and to a generalized
spherical function $P^\sigma_{mn}(\cos\theta)$ on the real 3-sphere.
Finally, on the surface of the real 2-sphere $S^2\sim\SO(3)/\SO(2)$
we have from (6) the usual spherical functions
$Y^m_\sigma(\cos\theta)$.
\begin{center}{REFERENCES}\end{center}{\small
\lit{1}{R. Takahashi} {Sur les repr\'{e}sentations unitaries des
groupes de Lorentz g\'{e}n\'{e}ralis\'{e}s //~Bull. Soc. math.
France. 1963. V.~91, P.~289--433.} \lit{2}{S. Str\"{o}m} {On the
decomposition of a unitary representation of (1+4) de Sitter group
with respect to representations of the Lorentz group //~Arkiv
f\"{o}r Fysik. 1969. V.~40, P.~1--33.} \lit{3}{V. V.
Varlamov}{Spherical functions on the de Sitter group //~J. Phys. A:
Math. Theor. 2007. V.~40, P.~163--201.}

\end{document}